# Responsible Adoption of Generative AI in Higher Education: Developing a "Points to Consider" Approach Based on Faculty Perspectives

Ravit Dotan
TechBetter, ravit@techbetter.ai

Lisa S. Parker
University of Pittsburgh, lisap@pitt.edu

John G. Radzilowicz
University of Pittsburgh, jgradz@pitt.edu

Written in collaboration with Members of the University of Pittsburgh

Ad Hoc Committee on Generative AI in Research and Education

**Abstract:** This paper proposes an approach to the responsible adoption of generative AI in higher education, employing a "points to consider" approach that is sensitive to the goals, values, and structural features of higher education. Higher education's ethos of collaborative faculty governance, pedagogical and research goals, and embrace of academic freedom conflict, the paper argues, with centralized top-down approaches to governing AI that are common in the private sector. The paper is based on a semester-long effort at the University of Pittsburgh which gathered and organized perspectives on generative AI in higher education through a collaborative, iterative, interdisciplinary process that included recurring group discussions, three standalone focus groups, and an informal survey. The paper presents insights drawn from this effort—that give rise to the "points to consider" approach the paper develops. These insights include the benefits and risks of potential uses of generative AI In higher education, as well as barriers to its adoption, and culminate in the six normative points to consider when adopting and governing generative AI in institutions of higher education.

## 1 INTRODUCTION

Over the last year, generative AI (GenAI) took academia by storm. The technology deeply impacts higher education on multiple levels, from the practicalities of day-to-day work to the ideals the sector seeks to promote [e.g., 24, 34, 76]. Given the magnitude of the current and potential future impacts of GenAI on higher education, institutions of higher education (IHEs) are pressed to decide how to approach the adoption of GenAI, and to make these decisions sooner rather than later.

With that in mind, many seek to better understand how GenAI impacts higher education and how to use and govern it responsibly in this context. Researchers often survey the perspectives of students, faculty, staff, and administrators, typically trying to understand how GenAI tools are used and key concerns regarding their use [e.g., 12, 14, 16, 19, 60, 65, 66]. Others conduct focus groups with students and faculty to explore the same questions, albeit less frequently [42, 61]. Yet others conduct desk analyses [e.g., 15, 76] or articulate their own views [e.g., 2, 69, 72, 75].

This paper presents insights and points to consider about GenAI in higher education arising from a semester-long effort at the University of Pittsburgh which included a semester-long semi-structured group discussion (29 participants), three standalone focus groups (19 participants overall), and an informal survey (144 respondents). Overall, the effort gathered and organized rich and in-depth perspectives drawn from individuals, particularly faculty members, working in the humanities, arts, sciences, social sciences, health sciences, and professional schools. **Section 2** describes the effort itself in more detail.

**Section 3** distills reasons for IHEs to adopt GenAI and barriers to adoption. In brief, the reasons for adoption point out that adoption is pivotal for the promotion of the goals and values of IHEs and their economic well-being. One challenge is the low faculty familiarity with both GenAI tools and their potential benefits and risks. Another challenge is the potential for the integration of GenAI to over-burden faculty in various disciplines.

**Section 4** presents GenAI use cases, benefits, and risks participants identified. These are especially valuable given the challenge of low familiarity, and they are complementary to the identification of uses, benefits, and risks articulated from other perspectives, such as intergovernmental agencies [74, 75] and students [12]. It also presents ideas for GenAI-related assignments that emerged in the focus groups. These include assignments that require the use of GenAI and assignments that decrease the potential for cheating using GenAI.

**Section 5** articulates points to consider for making decisions about GenAI in higher education, including decisions about where policy is needed, what to include in policies, and guidance for individual decisions regarding use of GenAI tools. It begins by examining the "top-down" approach taken by many private sector organizations, with management establishing rules that employees must follow. It explains why a different approach is often necessary in higher education due to the long-standing tradition of faculty governance of IHEs and academic freedom, a core value in academia. It then sketches an alternative for academia which recognizes that a centralized "top-down" approach is sometimes ethically and practically warranted, but that what is generally needed is another policy-making approach: "points to consider," points that should be taken into account when making decisions about GenAI use and policy-making. A "points to consider" approach is used in other domains that require room for individual judgment, context-sensitivity, and responsiveness to an unsettled regulatory landscape, such as decision-making by Institutional Review Boards (IRBs) [58]. The section presents six points to consider and discusses how they should be applied by being sensitive to features of the contexts in which GenAI may be used.

## 2 BACKGROUND

### 2.1 From Eliciting Interdisciplinary Perspectives to Presenting Points to Consider

The approach and specific recommendations presented in this paper are based on perspectives about GenAI articulated by individuals affiliated with the University of Pittsburgh during the Fall 2023 semester. The iterative process employed a stream of recurring group discussions, three standalone focus groups, and an informal survey that resulted in a corpus of interdisciplinary perspectives that were then organized and are presented in this paper.

The recurring group discussions took place as part of an ad hoc committee, formed at the request of the University's Provost and Senior Vice Chancellor (SVC) for Research. The committee's goal was to identify and report on topics where guidance is needed regarding GenAI applications in higher education. Twenty-nine committee members—with disciplinary backgrounds in the arts, humanities, sciences, health sciences, and social sciences—agreed to participate in the semester-long discussion. Most were faculty, but two members were from the Provost's office, six from the Chancellor's office, one from the University's community engagement office; one undergraduate and two graduate students were also members; and one member was an external consultant, specializing in AI ethics, who was previously affiliated with the University as a postdoctoral fellow (See the Appendix for the full list of committee members).

The committee met approximately every two weeks during the Fall semester, and it employed a normative, consensus-building process to organize concepts, clarify distinctions, and consider the interests and perspectives of stakeholders (gathered through the survey and focus groups). The committee began with unstructured conversations about GenAI opportunities and concerns in higher education resulting in a decision to divide the conversation into three topics: GenAI in research, teaching, and administrative work. The committee then dedicated meetings to each domain. The meetings elicited a full range of members' perspectives regarding risks and potential benefits of using (or not using) GenAI. Each time, the committee's insights were articulated in writing by Lisa Parker, one of the authors, who with the third author John Radzilowicz co-chaired the committee. Subsequent committee discussions resulted in emendation and elaboration of the text as the committee approached wide agreement on the "points to consider" that formed the normative core of the preliminary report submitted to the Provost and SVC for Research.

The standalone focus groups and informal survey were conducted as part of an initiative to map faculty's use and opinions about GenAI in eight units at the University of Pittsburgh: Business, Computer Science, English, Law, Physical Medicine and Rehabilitation, Physics, Psychology, and Theatre Arts.

The survey asked faculty in the above-mentioned units three questions: (i) whether they use GenAI tools in their teaching, research, service, grant-writing, or other professional activities (multiple selection); (ii) how many courses they were teaching in that semester; and (iii) how many of these courses had GenAI policies. The focus groups were open-ended conversations focusing on the following topics: (i) Do you currently use generative AI? How? (ii) Which uses of generative AI do you think are good, bad, or neutral? (iii) For the applications that you think are good, do you see any potential risks, downsides, or negative consequences? and (iv) What steps do you think should be taken to address the concerns? The focus group participants were faculty members in the same units we engaged in the survey. Data collection continued until data saturation was reached. Overall, 144 individuals from the above-mentioned units responded to the survey, and 19 individuals participated in the focus groups. Most of the focus group participants and survey respondents were tenured or tenure-stream, though some were outside the tenure-stream. The survey and focus group were led and analyzed by one of the authors, Ravit Dotan, who was also an active member of the Ad Hoc Committee.

The content of the paper is primarily based on the discussions of the committee. Some of the perspectives presented in this paper originated from the focus groups. We explicitly attribute these perspectives to the focus groups where that is the case.

### 2.2 Diversity of Perspectives and Limitations

The perspectives represented in this paper are diverse in important respects. The range of represented academic units is wide, with participants from the humanities, arts, social sciences, natural sciences, health sciences, and professional schools. The perspectives are also diverse in terms of gender distribution. About 54% of participants identified as men and 46% as women.

Having a diversity of disciplines and perspectives represented on the committee proved valuable in helping ensure that a full range of concerns, perceived benefits, and normative perspectives regarding GenAI was identified. A self-described engineer, for example, said he didn't understand what would constitute an "ethical framework" beyond risk-benefit analysis to minimize harms and maximize benefits. In response, an ethicist pointed to concerns about equity and to consideration of whether some interests are so important that they are protected by rights that must not be violated in seeking benefits. A self-described "non-traditional" student, who is "considerably older" than classmates and "Autistic and manag[ing] several challenging health conditions"—consistently reminded members of GenAI's potential to increase inclusivity, access, and equity for students with disabilities and cognitive differences, as well as first-generation college students and linguistic minorities, as it can be used to evaluate materials and online resources for appropriate elements of universal design. Debate about the degree of specificity of the recommended points to consider was similarly influenced by members' institutional positionality and interests: for example, some who were acutely aware of increased mental health needs of students and pressures on existing counseling services advocated for consideration of using AI/GenAI products in this domain, but other members were leery of recommending such a specific consideration, especially while the value of such tools was still being assessed. The committee ultimately advocated that the University "explore the use of GenAI/AI to support … university activities to create a holistically student-centered educational culture."

Nevertheless, the diversity of perspectives represented in this paper is limited in other important respects. First, the University of Pittsburgh is an R1 institution in a US urban setting. The circumstances of this type of institution are different from those of other types of IHEs, such as IHEs in other countries, IHEs in rural areas, liberal arts colleges, and two-year colleges. In addition, the vast majority of the participants were either faculty or administrators. Students and staff, whose positions in IHEs are very different, were a minority. Moreover, the racial and ethnic diversity represented in this effort was limited, with most of the participants being White. We do not have information about other important aspects of participant diversity, such as age, religion, and sexual orientation. Stakeholders that are external to the University of Pittsburgh and higher education, such as members of the municipal community and GenAI vendors, were not represented, although the University's community engagement office was represented.

We ask the reader to keep these limitations in mind when reading this paper. For a student perspective on GenAI in higher education in Hong Kong, see [12]. For an intergovernmental perspective on GenAI in higher education, see [74, 75]. For a survey of faculty and students from more than 600 higher education institutions, [see 19, 65]. For a survey of 404 higher-ed leaders about GenAI, [see 16].

### 3 GENAI IN THE CONTEXT OF HIGHER EDUCATION: REASONS FOR ADOPTION AND CHALLENGES

### 3.1 Reasons to adopt GenAI in Higher Education

#### 3.1.1 *The Advancement of Science.*
The values of accuracy, replicability, creativity, intellectual honesty, and integrity are prized in higher education because they are deemed integral to the pursuit of knowledge, which is at the core of the mission of higher education.

The incorporation of current GenAI tools in research activities threatens these values [6, 21]. For example, inaccuracy and bias in the output of GenAI are well-documented [e.g., 13, 26, 31, 51, 71], and the methods and algorithms by which the output is generated are opaque, which undermines the replicability of results [e.g., 7]. Opacity and bias threaten the integrity of the process [57]; 'intellectual honesty', a human virtue, is difficult to apply even as an analogy to a machine-learning model noted for bias [70].

Nevertheless, because IHEs are committed to pursuing high-quality research that benefits humanity, they should embrace using GenAI to improve research, conducting research on GenAI itself to increase its accuracy and potential benefits, and adopting policies and practices that limit the risks of employing GenAI. Just as IHEs, for example, adopt procedures to address the risks of "dual use research of concern"—life sciences research involving materials or methods that could be misapplied with a serious negative impact on humanity—IHEs might adopt measures that would minimize the risks and help ensure responsible use of GenAI [52, 64].

3.1.2 *Goals and Activities of Higher Education.*

Those in academia generally believe that acquiring skills of critical reasoning and a broad knowledge-base enables people to participate more fully and responsibly in public life as citizens of the world, members of their particular communities, and participants in the workforce and civic life. Therefore, in addition to seeking to expand knowledge, academics seek to impart their knowledge and the skills of its acquisition to their students and trainees.

Many of the activities central to higher education—e.g., analyzing and creating text—are affected by the introduction of GenAI, resulting in both opportunities and challenges. In short, GenAI tools may be used to supplement and enhance education, or to supplant or circumvent education.

Higher education has a responsibility to impart the knowledge and skills necessary to make informed decisions about the use and governance of GenAI. Therefore, it must also cultivate the knowledge-base and skills necessary to evaluate that use, which may require teaching students to proceed without employing GenAI prior to teaching them to use it, much as students are taught basic arithmetic before they are allowed and encouraged to use calculators.

Faculty developing curricula and courses must be equipped to make decisions about when it is appropriate to transition to the use of GenAI, i.e., when use of GenAI will enhance learning the underlying subject matter or when being able to use GenAI in a particular domain is itself a learning objective. They must also be equipped to recognize when GenAI may play a supportive role in equitably achieving traditional (pre-GenAI) educational objectives for a broad range of learners who may have different learning styles or different levels of previous background in the underlying subject matter.

3.1.3 *Economic Constraints and Competition.*

Higher education has been facing economic and competitive challenges for several decades [5, 29, 39]. Graduate programs, especially in the humanities and social sciences, have faced declining enrollment [8, 36, 37], and placement of graduates, even of STEM programs, into academic positions has been increasingly challenging for a variety of reasons [11, 62].

Declining birth rates and thus declining applicant pools affect higher education as a whole [39], although some elite and niche schools are less impacted [30]. The COVID-19 pandemic presented unprecedented challenges to institutions' viability [30, 38]. Institutions have competed with each other to offer amenities, provide distinguishing experiences (including experiential learning and intern/externships), and engage with the world outside academia [1, 44, 63].

Most recently, higher education has faced questions about the economic value or return on investment of a college education [10, 22, 27, 33, 73]. Increasingly, private industry is emerging as a competitor-educator of college-age, workforce-eligible young adults [17, 18, 23, 32, 59]. In light of the role that GenAI tools play in multiple industries [43], to remain competitive, it is imperative that institutions of higher education employ GenAI and educate their students about it and that research-focused institutions conduct research on it and its implications.

At the same time, there are practical and normative challenges to institutions of higher education entering the GenAI arena or even employing it in their usual activities.

## 3.2 Barriers to GenAI Adoption in Higher Education

3.2.1 *Currently Low Faculty Preparedness for GenAI Adoption and High Student Adoption.*

Faculty members' understanding and use of GenAI tools is generally low. In our informal survey, most faculty (63.9%) indicated that they don't use GenAI in their professional activities at all. The top use areas were research (21.5% of all respondents) and teaching (17.4%). Few faculty reported using GenAI in administrative work (6.9%). Moreover, only half of the courses taught by respondents in Fall 2023 had GenAI policies (50.5% of all courses taught by respondents). However, one of the departments, English, was an outlier. While more English courses (75.4%) had such policies, only 38.4% of courses outside of the English department had them. The low adoption and familiarity with GenAI tools were also reflected in our focus groups. In these, faculty strongly focused on GenAI in the context of addressing student use of the tools, such as in issues of academic integrity. Even when explicitly asked about the use of GenAI in research and service, faculty didn't have much to say.

The results of our informal survey are in line with formal surveys conducted by others. For example, a survey of 1000 faculty members across more than 600 institutions found that 75% of faculty members do not use GenAI regularly [19, 65]. At the same time, students are using GenAI. For example, the same survey also found that about 50% of the 1,600 students surveyed did use GenAI in their academic work [19, 65]. Another survey found that 84% of higher education leaders say that their institution is worried about AI-powered cheating [16].

Low adoption of GenAI by faculty presents challenges, especially in contrast with the high adoption among students. In our focus groups with faculty, the top concerns were related to this gap. The topics of most interest were academic integrity, re-thinking assignments in light of the availability of GenAI to students, and the impact of GenAI on the educational experience of students.

In particular, low preparedness for GenAI adoption may jeopardize the goals and values of higher education articulated above. First, GenAI has the potential to save time, and even contribute to scientific breakthroughs. Faculty members who are unfamiliar with the tools cannot take advantage of them. Second, insufficient familiarity with the limitations of the tools may lead to potentially problematic, even harmful, uses, such as using GenAI as a search engine without appreciating issues such as misinformation and bias or using GenAI to detect whether student papers are AI-generated (using tools that are unreliable and discriminatory [40]. Third, faculty members who are unfamiliar with how they can use GenAI in their own work are likely underprepared to teach their students how to employ the tools, teach them how to use them responsibly and think about their use and output critically, or appropriately identify and address GenAI–related academic dishonesty.

Given low familiarity with GenAI tools among faculty, efforts to identify potential uses as well as risks and benefits are key to progress. The next section presents the potential uses, risks, and potential benefits identified in the committee meetings and focus groups, including ideas raised for assignments incorporating GenAI or decreasing the possibility of using GenAI to cheat.

3.2.2 *Burdens Associated with Faculty Adoption.*

Faculty reported that incorporating GenAI into their professional activities would require substantial additional labor—for example, revising syllabi and creating new assignments are labor–intensive, as are identifying and evaluating GenAI tools. Faculty are concerned that this additional labor would be uncompensated and would be added to the expectations of people who are typically already over-extended and often under-compensated.

## 4 POTENTIAL USES OF GENAI IN IHES AND RESULTING RISKS AND BENEFITS

### 4.1 Identified Potential Uses of Generative AI

The committee discussions identified three domains for the potential use of generative AI. The uses listed here were identified as potentially beneficial. They are not necessarily recommended. Rather, they are uses to be considered in light of the points discussed in the subsequent sections. In particular, for reasons discussed in the next section, some of the potential uses were flagged as "sensitive uses" that would require heightened scrutiny and special consideration before adoption. Moreover, some uses are already subject to regulation by external bodies such as journals and publishers or research sponsors (e.g., see Nature's and Springer's policies on the use of generative AI in publishing and peer reviewing articles [67; 68], and the policy of the National Institute of Health (HIH) policy on the use of AI in peer reviewing grant applications [55]). For other perspectives on the use of generative higher education, see [12], [16], [19], [65], [74], and [75].

4.1.1 *Teaching and Learning.*

Instructors could incorporate GenAI into course and curriculum design; use it to generate assignments, as well as primary and supplemental teaching materials; communicate with students (e.g., creating chatbots to respond to frequent questions about material, assignments, or course expectations); incorporate elements of universal design to make materials accessible; and personalize educational experiences (e.g., customizing materials for multilingual English speakers or students with varying backgrounds, levels of preparation, or learning styles). Using GenAI to evaluate student work was deemed a "sensitive use" that requires special consideration, as described below. Moreover, IHEs, disciplines, and individual instructors also need to prepare students to use GenAI in their future careers. Concurrently, students could use GenAI tools as a personal tutor or coach for team learning, or to have material explained in multiple ways, to be asked open-ended questions, or to receive feedback [46, 47, 48, 49, 50].

In the focus groups, participants paid special attention to rethinking assignments in light of students' access to GenAI. Many were interested in designing assignments that would make GenAI-assisted cheating more difficult. Instructors brought up several ideas for such assignments, such as incorporating oral components into their courses, for example, by adding in-class presentations. In addition, asking students to generate mindmaps of arguments they are reading or presenting may help ensure that they are actively engaging with the material, particularly if they develop the mindmaps collaboratively with their peers. Further, students may be asked to annotate a text, e.g., to offer comments on an assigned reading. In collaborative annotation assignments, they respond to one another, build threads of commentary on the text, and may engage with the instructor. (One participant noted that tools such as Perusall can be helpful.) A common thread in annotations and mindmaps is that they do not focus on the production of a text. Instead, the student produces content in a structured way. Cheating by using GenAI is still possible, but would be less straightforward and may require more critical thinking even if GenAI is used. Instructors can keep this in mind to design other forms of structured assignments.

Discussion also considered how to change test conditions to eliminate or reduce GenAI-assisted cheating. Instructors may use tools that don't allow students to switch between tabs while working on the assignment (such as LockDown Browser). However, employing such tools may create accessibility problems for students with different learning styles or disabilities. While students could use other devices to access GenAI, their use would be more complicated and capable of being monitored or detected than when they simply access multiple tabs on one device.

Participants were also interested in discussing types of assignments that may effectively employ GenAI tools. These include:

- **Analyzing AI–generated output** – Here, the instructor presents the students with a GenAI generated output, such as a text, an image, a code, or a solution to a math problem. Students are asked to analyze this output. The task for the students could take different forms, for example:
- o **Identify themes** – The instructor generates text, code, a mathematical proof, or other outputs that illustrate themes discussed in class. Students are asked to identify these themes.
- o **Identify deficiencies** – The instructor generates text, code, a mathematical proof, or other outputs that illustrate mistakes or other deficiencies (including biases) in the GenAI-generated output that are relevant to the class content. Students are asked to identify the problems and fix them.

This type of assignment teaches students both about the course topic and about GenAI as a tool.

- **Revising AI–generated first drafts** – The students are asked to use GenAI to generate the first version of an assignment, such as a list of ideas, an outline for a paper, a summary of a paper, a piece of code, or an analysis of some data. Then, they are asked to improve on this first draft and to explain how they executed this improvement, including which prompts they used initially and how they improved the output either through further prompts or without the further use of GenAI. A goal is to reflect critically on the process of developing their output (e.g., paper, argument, or code) and the value of employing GenAI, as well as to learn how to write prompts that are suitable for their goals.
- **Interlocutor** – Students write text, code, a mathematical proof, or other output themselves. Then they ask GenAI to criticize their output. For example, students could assign to the GenAI different personas and ask for various critiques. Students then improve their initial product based on the interaction, describe the process, and critically reflect on it.
- **Generate practice questions** – Students who want additional practice could use GenAI to generate additional study questions. For example, they could feed into the GenAI old exam questions and ask for new questions in the same style. Another approach would be to feed an article into the GenAI and ask for questions about the article in the style of questions from old exams. Instructors can suggest best practices for using GenAI as a practice question generator. This type of assignment may be particularly helpful for students with different learning styles or those with less initial background than their classmates.

4.1.2 *Administrative and Service Activities.*

Committee work (e.g., generating minutes), aspects of pre-award and post-award sponsored research administration, and procurement (e.g., review of vendor bids and contracts) could employ GenAI. GenAI could provide drafts of communication with various internal and external stakeholders, including alumni and current or prospective students, or draft reports of activities for IHE leadership or press releases, as well as draft calls for proposals and applications. Faculty and staff could use GenAI to draft materials for their evaluation.

As described below, sensitive administrative uses of GenAI requiring special consideration would include its use in evaluating materials for student admissions, selection of postdoctoral or other trainees, or hiring and evaluating faculty and staff. Using GenAI to summarize or analyze students' evaluation of instructor teaching would require special consideration by instructors themselves and by administrators evaluating faculty.

4.1.3 *Research.*
IHEs conduct research on GenAI, including research on its implications and governance, and develop GenAI tools. Researchers could also use GenAI to search for, translate, and summarize information; analyze text and data; write computer code; draft or edit text (e.g., boilerplate parts of proposals or consent forms); draft reports of research for various audiences (e.g., public engagement regarding research or reports to funders, regulatory bodies, or IHE leadership). Relatedly, those in the creative arts—music, visual art, writing, film, and theater—could use GenAI as a primary or supplementary creative tool or to analyze creative work. GenAI could also be used in peer review of journal articles, book manuscripts, proposals of conference presentations, or research proposals, as well as by an IHE's research infrastructure (e.g., pre-award to ensure an application is complete, or post-award to identify conflicts of interest). Several research sponsors (e.g., the National Institutes of Health, National Science Foundation), journals, and publishers have issued guidance regarding use of GenAI in what is submitted to them for review [20, 35, 55, 56].

## 4.2 Risks, Concerns, and Potential Benefits Identified

Here we identify the perspectives articulated in the focus groups and committee meetings on the risks and potential benefits related to the adoption of generative AI in higher education (For other perspectives on such risks and benefits, see [12], [16], [19], [65], [74], and [75]).

Focus group and committee discussions identified two primary downside risks of not engaging in (responsible) use of GenAI: first, falling behind both peer institutions and companies (a) in attracting students and employees (both faculty and staff of all types) and (b) producing competitive graduates, research products, and creative output; and second, relatedly, not reaping the technology's potential benefits.

In these discussions, many potential benefits of using GenAI were identified. These included: advancing scientific progress, improving teaching and learning outcomes, reducing faculty and staff burden, realizing efficiencies, promoting equity, providing personalized disability-related assistance in education and employment, preparing students for the workforce, promoting informed citizenship, remaining competitive as an institution, and attracting and retaining faculty, staff, and students.

Nevertheless, many concerns were articulated, particularly with regard to the current state of the technology and the current level of faculty preparedness for its adoption. Chief among these were concerns about the inaccuracy of GenAI output and lack of expertise in evaluating GenAI outputs and using them responsibly. Many comments focused on concern about bias, including bias and gaps in the corpus or training data, bias in the output, exacerbation of bias and discrimination—either resulting from reliance on the output or existing inequities in access to GenAI and inequities in skill to employ it, and exacerbated marginalization of cultural and linguistic minorities and of neurodivergent students and other non–typical users. Loss of skills due to reliance on GenAI, loss of unique voice in written products, and displacement of labor and expertise were other concerns articulated. Homogenization of voice and loss of skills were identified as special concerns in higher education where goals include nurturing creativity and curiosity, development of critical reasoning skills and specialized disciplinary skills, and development of both perspective-taking skills and students' own worldviews. Because development of such skills is iterative and progressive, concern focused on students' using GenAI to skip development foundational skills and deep understanding of the concepts on which they rely. Concern also focused on how GenAI output could reflect and amplify "echo chambers" of more prevalent views to the exclusion of unusual or minority perspectives, and how embracing GenAI could reinforce the acceptability of this "majority data rule" approach in analyzing questions, texts, and the corpus of higher education.

Unsettled intellectual property (IP) and copyright issues, privacy and confidentiality infringements, and impact on the creative arts and artists were cited. Increased faculty and staff burden— including to learn to use, or to police the use of, GenAI was of concern, as was the potential to overwhelm an IHE's research administration infrastructure. Legal and reputational risks (to the institution and to GenAI users, both units and individuals) were noted, as was concern about undermining academic integrity, supporting exploitative training of AI tools (i.e., use of people's work as training data without compensating them), and the negative environmental impact of GenAI. With regard to research, there was concern about the difficulty of accurately estimating the cost of employing GenAI (and AI in general) in research projects, and thus the risk to an IHE of being committed to "eating" cost overages in order to fulfill obligations of research grants and contracts awarded to it.

# 5 AN ALTERNATIVE TO CENTRALIZED, TOP-DOWN POLICY-MAKING: POINTS TO CONSIDER REGARDING GENERATIVE AI IN HIGHER EDUCATION

## 5.1 Tensions between Centralized Policy-making and Academia's Structural Features, Values, and Goals

Approaches to AI responsibility, and GenAI responsibility in particular, that are common in the industry utilize a "top-down" approach, where management decides on policies or rules to which the organization's employees are expected to adhere [e.g., 9, 45, 28].

This top-down approach is at odds with the shared governance or faculty governance structures of IHEs, as well as the ethical norm of academic freedom that emerges from such self-governance and that is integral to IHEs [41]. According to this norm of academic freedom, faculty have the freedom to express ideas—in teaching and in their research and scholarship—without fear of negative sanctions (e.g., loss of employment or opportunities for advancement, censorship, repression, or prosecution). Individual faculty members and researchers generally have the right to choose their question and methods of inquiry, as well as the methods and requirements they wish to employ in their classes, albeit within the curricular parameters established by the disciplines and units in which they are employed. This freedom of inquiry and the concomitant absence of control by the state or other authorities are considered necessary for the pursuit of knowledge and the advancement of science. The freedom to allow "a thousand flowers to bloom," or at least encouragement of intellectual creativity as opposed to lockstep compliance with prevailing norms, is integral to the increase of knowledge.

The advancement of science and pursuit of truth and knowledge (research)—and preparing others to do so (education)—are the raison d'être of IHEs. Thus, adopting a centralized or "top-down" approach to employing and governing GenAI would be contrary to the norm of academic freedom and the shared-governance ethos of academia, as well as contrary to the intellectual or scientific desirability of encouraging multiple approaches to the use of GenAI in research and education.

Moreover, in contrast to companies that are typically structured to be nimble in responding to identified social needs and market opportunities, educational institutions may place greater emphasis on involving multiple stakeholders in decision-making processes, have necessarily slower processes of decision and policy-making (e.g., to implement curriculum changes), and often have to appeal to state legislators and boards of trustees to make changes that then may take several years to implement and market to potential students and employees. For these practical, structural reasons, leaders of IHEs can seldom act as swiftly or quasi-unilaterally as industry CEOs.

On the other hand, some academic and ethical values may at times be best promoted by embracing a top-down approach. In procurement, for example, not only might there be economies of scale when IHEs negotiate with vendors of GenAI tools, rather than leaving negotiation and purchase to individuals or individual units (e.g., departments, programs, or laboratories), but also negotiation by the institution as a whole can help promote equity and ensure that units of differing size with differing resources have equitable access to GenAI and to expertise to employ it responsibly. Further, because different vendors have different policies with regard to, for example, the ownership and protection of the privacy of data fed into GenAI tools, IHEs are likely to have the authority—and access to the expertise—to investigate those vendor policies to help ensure responsible protection and use of that data. IHEs are also more likely than individuals to be able to require vendors to reveal data about bias in their products' training data and the risk of inaccuracy and bias in their output.

Similarly, it is ethically appropriate for an IHE to adopt a uniform academic integrity policy regarding the use of GenAI. While different instructors may have some latitude regarding what uses of GenAI are permissible in their individual classes, it is appropriate for an IHE to promulgate guidelines, or parameters within which variation is permissible, so that its students are ensured transparency and some degree of stability regarding expectations of GenAI use. Therefore, the responsible introduction of GenAI in academic contexts may require an innovative approach, which combines "top-down" components within the confines of the embrace of institutional shared governance and academic freedom. Guidance, such as points to consider or the results of centralized vetting of GenAI products, may be provided while respecting various domains reserved for personal judgment or unit-level policies.

Because of these tensions, the committee recognized that in some domains in higher education, a more centralized approach to GenAI policy-making is appropriate—namely, academic integrity, the procurement of GenAI tools, and the use of GenAI in sensitive processes such as hiring and admissions. However, more generally, the committee encouraged thoughtfulness about when centralized policies are appropriate.

Instead of adopting a top-down, centralized policy approach, the committee identified points that should be considered when deciding whether and how to establish policies about GenAI and when individuals decide whether and how to use GenAI when they are not constrained by a policy. This approach, of providing substantive normative points that together provide a framework for "thinking through" and identifying ethically justifiable courses of action, respects faculty members' academic freedom to exercise their judgment while providing guidance for that judgment.

The "Points to Consider" approach provides a framework within which to consider relevant issues rather than providing strict rules. The approach is widely used in domains characterized by context-dependency, unsettled legal and regulatory consensus (particularly regarding emerging technologies), or the need to maintain room for individual judgment. The points to consider approach has been employed, for example, by the Food and Drug Administration (e.g., regarding characterization of cell lines to produce biologicals, or testing of monoclonal antibody products), the National Institutes of Health (e.g., regarding research with individuals who have questionable decisional capacity), the Recombinant DNA Advisory Committee (regarding gene therapy), and the pharmaceutical industry [25, 53, 54], as well as in research ethics consultation [4]. The Office of Research Protections, for example, used the Points to Consider approach throughout its 1993 *IRB Guidebook* to respect the decisional authority of local institutional review boards while providing some guidance for their exercise of judgment [58].

## 5.2  Points to Consider Grounded in Ethical Values and the Values of Higher Education

In its discussions, the committee identified four values-based points to consider. First, it determined that, as discussed above, **respect for academic freedom,** a fundamental value in higher education, serves as a constraint on institutional policy, guidance, and action regarding GenAI. Faculty should have latitude in deciding whether and how to adopt GenAI tools in their research and teaching.

Second, the integration of GenAI into higher education should **be consistent with academic and scientific values**, such as accuracy, replicability, creativity, intellectual honesty, and integrity. Therefore, when employing GenAI or making policy about its use, it is important to ensure that scientific standards are not compromised. Use of GenAI must be transparent—i.e., its use should be disclosed to support intellectual honesty and academic integrity, and how it is used should be explained in relevant detail in order to ensure accountability and facilitate replicability.

Third, use of GenAI and the development of guidance regarding it should seek to **minimize risk** of various kinds of harm, including discrimination, misinformation, physical and mental harm, and reputational harm. Therefore, for example, institutional policies and educational offerings regarding GenAI should specify best practices for risk mitigation, thereby encouraging their adoption. Moreover, given the potential for the use of GenAI to have a far-reaching material impact, it may be appropriate to consider developing a policy and procedures modeled on policies governing dual-use research of concern that provide additional review, oversight, and monitoring of the use of GenAI (and AI more generally). The risks associated with some uses or use contexts are especially great. These include the uses, described above, that the committee identified as sensitive.

Fourth, the use of GenAI tools has the potential to both exacerbate and mitigate inequities. GenAI use in IHEs should **seek to mitigate inequities** both in access to GenAI and through the use of GenAI and its output. The introduction of GenAI into higher education should not result in "rich departments getting richer," or student success becoming increasingly dependent on financial resources or prior familiarity with GenAI. IHEs should identify ways to ensure that students, faculty, and units (e.g., departments) have equitable access to GenAI resources to meet their differential needs, including opportunities to become "literate" with regard to the technology itself. Professional development, particularly for staff, should include opportunities to develop such GenAI literacy.

Moreover, IHEs may play a critical role in illuminating the social implications of employing GenAI and using its output, including its biases and potential to exacerbate inequities. Interdisciplinary research may explore the social implications of using GenAI, including potential inequities, and then articulate and test ways to mitigate them. IHEs are uniquely situated to engage in such interdisciplinary exploration, as they have the depth of expertise in the humanities and social sciences that industry often lacks.

## 5.3  Pragmatic Points to Consider for GenAI Policy Development in Higher Education

Two additional considerations were considered of practical importance even if they carry less ethical weight.

To **reduce the regulatory burden** due to new rule-making, existing IHE rules, policies, and guidance documents may be amended to address GenAI or may already be applicable to it. Students may need specific guidance regarding use of GenAI in their courses, labs, research, writing (e.g., of text, code, or music), and other creative endeavors. Often, student-focused guidance might be provided by adapting or revising existing IHE policies, while the use of GenAI during intern/externships, experiential learning outside the IHE, and study abroad experiences may necessitate students becoming familiar with and adhering to policies regarding GenAI of companies and other entities, including geopolitical entities.

For many contexts of GenAI use, IHEs may rely on existing and emerging rules and guidance regarding GenAI issued by journals, publishers, and research sponsors, as well as professional societies and governmental agencies. To respect academic freedom and foster shared governance, IHEs may choose, for example, not to develop policy regarding faculty use of GenAI in writing manuscripts or research proposals, and instead rely on the policies of external entities and the judgment of faculty, researchers, and scholars. IHEs might choose to provide guidance or points to consider to inform that judgment, given that many faculty are unfamiliar with GenAI tools and their limitations and that IHEs have an interest in protecting themselves and members of their communities from material and reputational risks.

In general, IHEs should take an educational approach, rather than a regulatory one, to increase awareness of the potential uses and benefits, risks, and limitations of GenAI. Faculty should be equipped to justify prohibiting, limiting, permitting, or requiring use of GenAI in educational contexts. Faculty, staff, and students should have educational resources to support their own decisions and responsible use of GenAI.

However, IHEs may sometimes need to develop rules or policies. For practical and ethical reasons, when IHEs themselves act as research sponsors (e.g., for internal grant competitions) or publishers (e.g., of a student journal), for example, they may need to adopt policies regarding use of GenAI to create or review work. In addition, if the use of GenAI were to increase the volume of research applications to be processed by an IHE's office of sponsored research, for example, new rules or procedures might have to be implemented to manage the increased workload.

Finally, rules, policies, and guidance regarding GenAI should be sufficiently broad and adaptable to **maintain adaptability to rapid change of GenAI tools.** They might be "time stamped" for future review to ensure their continued applicability. IHEs might create and charge interdisciplinary bodies with such ongoing review. The interdisciplinary nature of such bodies is important because different disciplines and different contexts of GenAI use may have different needs and present different considerations for the responsible use of GenAI.

5.4 **Applying the Points to Consider: Trade-offs between Features of the Context**

As is evident in other domains of ethical reasoning, those developing policy or considering specific instances of potential GenAI use cannot answer questions of whether and how to use GenAI by considering the aforementioned "points to consider" alone. In bioethics, for example, four principles (autonomy, beneficence, nonmaleficence, and justice) must be applied–including typically giving greater weight to one or another principle–in specific contexts involving either to develop policy or to resolve specific conflicts or make individual decisions [3]. So, for example, in developing bioethical policy (e.g., about aid in dying) contextual features like cultural, socioeconomic, and political factors matter. At the bedside, a patient's own values and preferences provide a critical context for decision-making involving the application of bioethical principles to address the individual patient's situation.

With regard to GenAI, the committee identified several contextual features that are relevant to the justifiability of using GenAI in both particular types of cases and specific instances. Typically, it is the interaction of features of the context that increases or decreases the level of concern appropriate to a particular use of GenAI, as there are trade-offs to be made in light of the relative importance of these features.

Goals.

Use of GenAI should be justified in virtue of its use serving the goals of the activity, and each step of the use of GenAI must be justified. Many activities in higher education involve writing text, for example; however, the goals of writing text vary widely depending on factors such as the context, audience, and writer of the text.

In educational contexts, decisions about whether and how to incorporate GenAI in courses (and particular components of the course like assignments or examinations) depend on the pedagogical goals of the course (and its components). Faculty should explicitly consider what they are trying to teach their students and whether having students use GenAI would inhibit or enhance that. Students may be asked to write text, for example, to demonstrate their understanding of the subject matter (e.g., on an examination) or to clarify (to themselves, for their personal benefit) their thoughts or emotions. The goals of writing in those contexts may not be served by employing GenAI. Using GenAI to polish text might be helpful to a multilingual student whose English grammar skills are still developing when she is asked to report the findings of a laboratory experiment; however, continual reliance on GenAI could prevent the student from developing those language skills herself. Moreover, the student would need to have sufficient adequate language skills to verify the appropriateness of the GenAI output.

Researchers might use GenAI to draft a description of their studies in "lay language" for a press release, or for inclusion in consent forms, which should have a reading level no higher than 8th grade. GenAI could save time when drafting boilerplate portions of grant proposals; however, research trainees need to learn how to draft such boilerplate, so that they can both do so themselves and verify the accuracy and relevance of what GenAI produces.

Treating sections about nonhuman animal welfare as boilerplate that may be left to GenAI to draft may circumvent a goal of Institutional Animal Care and Use Committees that want researchers to think carefully about the number of animals needed to have statistically significant results and the procedures they will employ to minimize animals' suffering.

GenAI may be used to write computer code. In some more introductory classes, learning to write code is a goal. Relying on GenAI to complete assignments would circumvent the goal of learning how code is written. In more advanced classes, however, students might not only have developed the skills to verify the output of GenAI, but also have different goals for writing code—for example, to write and use code to analyze data. In those advanced course contexts, employing GenAI to write code could serve course goals and supplement learning. In advanced studio arts classes, GenAI might be used as a medium of creation, while in an introductory class, its use might circumvent development—or demonstration of development—of basic concepts and techniques (e.g., perspective and sketching).

### 5.4.1 *Material Impact.*

The risks involved in the activity and the importance of the action to be taken on the basis of the activity must be considered when deciding whether and how to use GenAI. Moreover, material impact has different dimensions, including its immediacy, reversibility, and magnitude of importance for individual and/or group well–being.

The material impact of various activities in higher education differs; the material impact of employing GenAI differs accordingly. Learning exercises, admissions or employment decisions, and generating text on the basis of which diagnosis or medication prescription will be made, for example, have different levels of material impact in terms of the immediacy, probability, magnitude, and reversibility of material impact. They also differ with regard to the presence or absence of oversight or a system of checks-and balances that might discover and reverse or mitigate negative impact. Individuals considering use of GenAI—and institutions of higher education developing guidance and policy regarding its use—need to consider how actions taken based on activities including GenAI output could affect the well–being of individuals, groups and communities, and the environment.

A student might use GenAI, for example, to write text or code prior to developing requisite skills (or adequate knowledge of the underlying subject matter) herself. She may have good luck and receive output that is sufficiently accurate, even though she may lack the skills to verify that fact. She may receive a good grade on the assignment and in the class. Eventually during her course of study, her gap in learning may undermine her academic success; indeed, it would be fortunate for a future assignment or course to identify and prompt her to correct this lacuna. The progressive nature of a curriculum may be considered a "check-and-balance" to ensure students' adequate mastery of skills and bodies of knowledge. In the future, in the "real world," the student may not be required to use the skill or knowledge she was to have learned-she might pursue a different career path—and thus the impact on others of her inappropriate reliance on GenAI would have little material impact on others. Or, she might "slip through the cracks" of the curriculum and her inadequate preparation could affect others, as well as her long-term career. Ideally, the greater the risk of material harm from an inadequately prepared professional, however, the more likely is the presence of measures to ferret out that inadequacy (e.g., licensure examinations, multiple review of work products like review of architectural plans). In some contexts, such systems of oversight pre-exist GenAI, and may be adapted or increased with its implementation in various fields. Other contexts lack such oversight mechanisms, or their adequacy is unclear.

The immediacy and magnitude of impact, as well as the lack of transparency and irreversibility, of decisions to hire an employee based on output from GenAI evaluating application materials led the committee to designate as a "sensitive use "the use of GenAI in evaluating, hiring, promoting, or firing employees. Some committee members suggest not using GenAI in any aspect of such human resources activities; others would limit its use to verifying basic completeness of application or annual evaluation materials, drafting boilerplate communications (e.g., "Thank you for your application …), or drafting job announcements. By designating such uses as sensitive, the committee indicated that use of GenAI in those contexts warrants special consideration or heightened scrutiny prior to its adoption, requires more constant oversight if it is to be used, or might be altogether prohibited because of the risks involved.

In addition to its use in human resources decision-making—including hiring, evaluation, retention, and promotion of faculty, as well as staff—the committee considered as similarly sensitive the evaluation of materials for student admissions or selection of postdoctoral or other trainees, as well as evaluation of student work for grades and for awards of degrees or honors. Using GenAI to summarize or analyze students' evaluation of instructor teaching would similarly require special consideration by instructors themselves and by administrators evaluating faculty. All of these activities have immediate and potentially irreversible impact on individuals, and the decision-making processes (e.g., grading, evaluating a resume or dossier) are not transparent to those affected. It is unclear what oversight could be implemented to adequately guard against the effects of bias and inaccuracy in the output of GenAI.

5.4.2 *Adequacy and Knowledge of Relevant Guardrails, Safety Measures, Policies, and Rules.*
The committee's discussions sought to identify whether there were already adequate guardrails, policies, rules, and guidance to ensure the responsible use of GenAI. It agreed that those with the expertise and authority to establish rules, policies, guidelines, and best practices should consider establishing these safety measures or adapting existing policies and guidelines to include GenAI use. These measures may be implemented at different levels (e.g., the user, particular use, particular type of use, department, laboratory, or institution). Determination of which level is most appropriate must be attentive to the various considerations the committee identified, especially respect for the role of faculty in the institution's shared governance and for academic freedom, concern about material impact, and concern for equity.

The committee also recognized that given the rapidly evolving landscape of GenAI development, the question of whether adequate safety measures exist is a question to be asked constantly by those considering use of GenAI. The institutional climate and guidance should prompt potential users of GenAI to ask whether such measures exist and whether they are adequate given the potential material impact of GenAI use in the particular context.

Thus, users should ensure that they are aware of the relevant rules, policies, guidelines, and other normative guidance prior to employing GenAI tools (such as [74], [75], [67], [68], [55]). However, they should also adjust their use of GenAI to the adequacy of the guardrails in place, given the potential material impact of the application. If there is no guidance—i.e., no guardrails, policies, and rules to which to adhere, users should carefully consider whether to employ GenAI at all. They might, for example, consult with colleagues who have no conflicts of interest regarding its use in the context at hand.

Users should also consider the conditions under which the output of GenAI would be used. When the output is to be used without or with only minimal human review, plans for its use warrant rigorous scrutiny. Were a GenAI application to be employed with minimal human oversight prior to its output having material effect, the stakes involved should be low (e.g., employing a chatbot to answer a basic inquiry may warrant little concern, while evaluating an employment application and rejecting an applicant should be considered a sensitive use and should be subject to rigorous scrutiny, if such "autonomous" GenAI is to be used in that context at all).

5.4.3 *Knowledge about the Tool.*
As the potential material impact of relying on GenAI and its output increases, so does the warrant for the user to understand the GenAI tool. Such understanding would include knowing how the tool works and its limitations. At least in broad strokes, potential users need to understand how the tool works, e.g., is the tool connected to the internet? To what kind of information does it have access? Who will have access to the information the user inputs, who will own it, and what may be done with that information? Are there appropriate protections of the security, privacy, and confidentiality of the information input? Does use of the GenAI tool—or did its use of training data—infringe protections of IP or copyright? The difficulty of individual users ascertaining information to address some of these questions is one reason that it is likely appropriate for IHEs to vet GenAI tools and make available to students, faculty, and staff tools that meet appropriate standards regarding ownership and use of data input, security and privacy protections, and protection of IP and copyright.

Users must also know the tool's limitations, for example regarding accuracy and bias of its output. Users must thus have adequate understanding of the potential bias and gaps in the corpus or training data of the GenAI tool proposed for use. They must also know how to produce and refine meaningful results. Institutions of higher education should educate their faculty, staff, and students regarding how to craft meaningful GenAI prompts and the iterative process of using such prompts.

Education for students in the responsible and meaningful use of GenAI may be incorporated into curricula and courses, and may also be provided in standalone short courses and workshops. Some education will need to be discipline-specific, but some may be more generally about the use and limitations of GenAI.

Faculty similarly need educational opportunities—both general and discipline-specific; these may be provided through in-house workshops and short courses, or faculty may be incentivized to seek education through their professional societies and organizations. Professional development activities for staff members should include opportunities to learn about GenAI use that is not only relevant to their professional work, but also pertinent to their roles as informed citizens. The opportunity to learn about this technology and its potential benefits, risks, and limitations may be considered a benefit of being employed in higher education.

5.4.4 *Knowledge about the Subject Matter.*

It is critical that the user be able—i.e., be qualified and have the opportunity—to review the output of GenAI to identify and fix inaccuracies, biases, and other problems. To be able to recognize and redress errors and biases, users must have adequate knowledge of the subject matter on which the tool is employed. When using GenAI to summarize a text about a topic with which one is unfamiliar, for example, one might not be able to identify even gross inaccuracies. A faculty member using GenAI to produce the first draft of a syllabus needs to be able to identify discriminatory omissions (e.g., ignoring key texts written by authors from backgrounds underrepresented in academia or in a literary field). If the faculty member is not sufficiently knowledgeable to identify and remediate such unintended bias, the use of GenAI risks perpetuating and exacerbating it.

As the potential material impact of relying on GenAI increases, so does the need to be able to evaluate its outputs. Therefore, in many contexts, novices should not be allowed to use GenAI in the same way as advanced individuals. As discussed above, the goal of higher education to develop skills and impart knowledge also supports limiting some people's access to GenAI until they have adequately mastered the skills and knowledge base that its use may supplant.

## 6 CONCLUSION

This paper proposes an approach to the responsible adoption of GenAI in higher education based on an in-depth iterative process involving University personnel, especially faculty. Based on insights gleaned from focus groups, a survey, and a semester-long semi-structured interdisciplinary discussion, the paper argues that the top-down approach to AI management prevalent in the commercial sector, in which senior leadership makes centralized decisions for the organization and its members, is a poor fit for the higher education sector. Instead, we argue for utilizing a "points to consider" approach, which leaves ample room for individual decision-making and the shared faculty governance traditional in IHEs. This approach is, we argue, more compatible with the goals, values, and structural features of higher education. We articulate six key points to consider when adopting and governing the use of GenAI in higher education, and explore the trade-offs involved when applying those points to evaluate particular types of use of GenAI and specific instances of its uses.

## 7 RESEARCH ETHICS AND SOCIAL IMPACT STATEMENT

### 7.1 Ethical Considerations Statement

This paper relies on discussions among groups of individuals and a survey. These activities involve risks that are common to qualitative research, such as revealing information participants have shared in confidence. In light of these risks, we consulted with our Institutional Review Board (IRB) before the beginning of the interactions. The IRB determined that the activities reported here are not Human Subjects Research subject to the review and oversight of the IRB, and thus were exempted from IRB review. However, to protect participants, we used the following precautions: (1) the survey was anonymous; (2) no information about the identity of the focus groups was shared with anyone except those who were there and the organizer; (3) before deciding to publish this analysis publicly, the ad hoc committee that engaged in the semester-long discussion voted unanimously to share the insights, conclusions, and points to consider contained herein.

### 7.2 Research Positionality Statement

The 29 committee members formed an interdisciplinary group, primarily of faculty (including 6 with high-level administrative or leadership roles), but also including 4 students and trainees. Some members had substantial previous experience with AI/GenAI, while others did not. The group represented a range of previous experience with policy-making and/or with identifying issues for policy-making. In addition to the 29 committee members, the paper conveys the additional 19 participants of the focus groups. One of the reasons for conducting the focus groups is that we wanted to explore the use of GenAI in disciplines not represented among the committee members while still maintaining a group size conducive to meaningful iterative discussion.

The range of disciplines represented, with a concomitant diversity in research programs and methods, as well as teaching modalities and styles, was considered an asset to the goal of identifying uses of GenAI in IHEs, as well as potential benefits, risks, and barriers to its use. We recognize, however, that the perspectives of people not affiliated with the University were filtered through a faculty member who directs the University's office of community engagement. In future discussions, incorporating more direct input from other stakeholders—community and municipal members, developers and vendors of GenAI tools, and a diverse range of students —would enrich the analysis. Because this paper results from inquiries and discussions at one RI institution, it reflects a single institutional culture. It is hoped that this presentation of the "points to consider" approach and the specific points proposed will inspire differently situated IHEs (e.g., two-year colleagues, liberal arts colleges, rural institutions, and non-US IHEs) to undertake their own effort to explore the points to consider approach, and the specific points to consider discussed in this paper. In doing so, they may identify additional points to consider that would help others or challenge the points we have raised.

Several of the committee members, particularly those from the University's teaching and learning center, addressed the potential benefits of GenAI use for those with disabilities and for multilingual students and researchers; however, only two of the members specifically identified as members of those stakeholder groups. Further, while the comittee discussed bias in GenAI training data and outputs, as well as issues of equity in access and risks of inequities being exacerbated by both the use and nonuse of GenAI, future discussions might benefit from more focused attention to the views of those self-identifying as speaking from different cultural, ethnic, and socioeconomic backgrounds.

### 7.3 Adverse Impact Statement

Advocating a "points to consider" approach and recommending the six key points elucidated herein is designed to minimize potential adverse impacts either for those who participated in the effort reported, or for those who adopt the approach we recommend, because it encourages decision makers to carefully think through their decisions. Having said that, the "points to consider" approach could risk adverse impact if it were abused. Misuses of the approach could include asserting the right to exercise individual judgement in order to avoid accountability, to flout ethical norms, or to create a mere façade of carefully considered decision making.

# 8 APPENDIX

Below is the full list of the members of the University of Pittsburgh Ad Hoc Committee on Generative AI in Research and Education:

- Keith Caldwell, EdD (education)

Executive Director of Place Based Initiatives

Office of Engagement & Community Affairs

- Michael Colaresi, PhD (political science)

Associate Vice Provost for Data Science

William S. Dietrich II Professor of Political Science

- Robert K. Cunningham, PhD (cognitive and neural systems)

Vice Chancellor for Research Infrastructure

Research Professor, Department of Informatics and Networked Systems

- Ravit Dotan, PhD (philosophy)

Data Technology Ethics Consultant

Former Director of The Collaborative AI Responsibility Lab, Center for Governance and Markets

- April Dukes, PhD (neuroscience)

Faculty and Future Faculty Program Director, Engineering Education Research Center

Swanson School of Engineering
- Bonnie Falcione, PharmD (pharmacy)

Associate Professor of Pharmacy and Therapeutics
- April Flynn, MFA (writing)

Teaching Associate Professor
Composition Program, Department of English
- Na–Rae Han, PhD (linguistics)

Teaching Professor of Linguistics
Director, Robert Henderson Language Media Center
- Michael Holland, PhD (analytical chemistry)

Vice Chancellor for Science Policy and Research Strategies
- Jennifer Iriti, PhD (developmental and educational psychology)

Assistant Vice Chancellor for Research Inclusion and Outreach Strategy
Research Scientist, Learning, Research & Development Center
- Robin Kear, MLIS (library and information science)

Liaison librarian in the ULS Research and Educational Support Department
Faculty Assembly President
- Alan Lesgold, PhD (psychology)

Emeritus Dean and Professor
School of Education
- Diane Litman, PhD (computer science)

Professor of Computer Science
Senior Scientist, Learning Research and Development Center
- Michael Madison, JD (law)

Professor of Law
Senior Scholar, Institute for Cyber Law, Policy, and Security
- Nora Mattern, PhD (library and information science)

Teaching Assistant Professor of Computing and Information
Director, Sara Fine Institute
- Ian Neumaier (majors: philosophy and law, criminal justice, and society)

Undergraduate in the Frederick Honors College
- Lisa S. Parker (philosophy)

Dickie, McCamey & Chilcote Professor of Bioethics
Director, Center for Bioethics & Health
Director, Pitt Research's Research, Ethics and Society Initiative (RESI)
- Clyde Wilson Pickett, EdD (education)

Vice Chancellor for Equity, Diversity, & Inclusion
- Lara Putnam, PhD (history)

Professor of History
Director, Global Studies Center
- John G. Radzilowicz, EdD (instruction and learning)

Director, Pedagogy, Practice, & Assessment
University Center for Teaching and Learning
- Matthew Roberts, MEd (education)

EdD Candidate, School of Education
- Jennifer Seng, JD (law)

Assistant Vice Chancellor and Deputy Chief Legal Officer, Office of University Counsel
- John Stoner, PhD (history)

Teaching Professor of History
- John Wallace, PhD (sociology)

Vice Provost for Faculty Diversity and Development
David E. Epperson Chair and Professor, School of Social Work
- David Wert, PhD (rehabilitation science), MPT (physical therapy)

Vice Chair of Doctor of Physical Therapy Education

Associate Professor, School of Health and Rehabilitation Sciences
- Katherine Wood, PhD (molecular and cellular physiology)

Research Assistant Professor of Pulmonary, Allergy and Critical Care Medicine
- Jennifer Woodward, PhD (microbiology and immunology)

Vice Chancellor for Sponsored Programs and Research Operations
Professor of Surgery and Immunology
- Shandong Wu, PhD (computer science)

Associate Professor of Radiology, Biomedical Informatics, and Bioengineering
- Joseph Yun, PhD (informatics)

Research Professor of Electrical and Computer Engineering